\documentclass[fleqn,usenatbib]{mnras}

\usepackage{savesym}
\usepackage{amsmath}
\savesymbol{iint}
\usepackage{txfonts}
\restoresymbol{TXF}{iint}

\usepackage[T1]{fontenc}
\usepackage{ae,aecompl}

\usepackage{graphicx}	
\usepackage{amsmath}	
\usepackage{amssymb}	
\usepackage{natbib}
\usepackage{subcaption,tikz}
\tikzset{boximg/.style={remember picture,red,thick,draw,inner sep=0pt,outer sep=0pt}}

\newcommand{\edit}[1]{#1}

\title[LMC deflecting the HVS]{Deflection of the hypervelocity stars by the dance of the Milky Way and Large Magellanic Cloud}

\author[D. Boubert et al.]{
D. Boubert,$^{1}$\thanks{E-mail: douglas.boubert@magd.ox.ac.uk}
D. Erkal$^{2}$
and A. Gualandris$^{2}$
\\
$^{1}$Magdalen College, University of Oxford, High Street, Oxford OX1 4AU, UK\\
$^{2}$Department of Physics, Faculty of Engineering and Physical Sciences, University of Surrey, Guildford GU2 7XH, UK\\
}

\date{Accepted XXX. Received YYY; in original form ZZZ}

\pubyear{2019}

\begin{document}
\label{firstpage}
\pagerange{\pageref{firstpage}--\pageref{lastpage}}
\maketitle

\begin{abstract}
Stars slingshotted by the supermassive black hole at the Galactic centre will escape the Milky Way so quickly that their trajectories will be almost straight lines. Previous works have shown how these `hypervelocity stars' are subsequently deflected by the gravitational field of the Milky Way and the Large Magellanic Cloud (LMC), but have neglected to account for the reflex motion of the Milky Way in response to the fly by of the LMC. A consequence of this motion is that the hypervelocity stars we see on the outskirts of the Milky Way today were ejected from where the Milky Way centre was hundreds of millions of years ago. This change in perspective causes large apparent deflections in the trajectories of the hypervelocity stars, which are of the same order as the deflections caused by the gravitational force of the Milky Way and LMC. We quantify these deflections by simulating the production of hypervelocity stars in an isolated Milky Way (with a spherical or flattened dark matter halo), in a fixed-in-place Milky Way with a passing LMC, and in a Milky Way which responds to the passage of the LMC. The proper motion precision necessary to measure these deflections will be possible with the combination of \textit{Gaia} with the proposed \textit{Gaia}NIR successor mission, and these measurements will unlock the hypervelocity stars as probes of the shape of the Milky Way, the mass of the LMC, and of the dance of these two galaxies.
\end{abstract}




\section{Introduction}
\label{sec:introduction}

\begin{figure*}
    \centering
	\includegraphics[scale=0.8,trim = 5mm 40mm 20mm 3mm, clip]{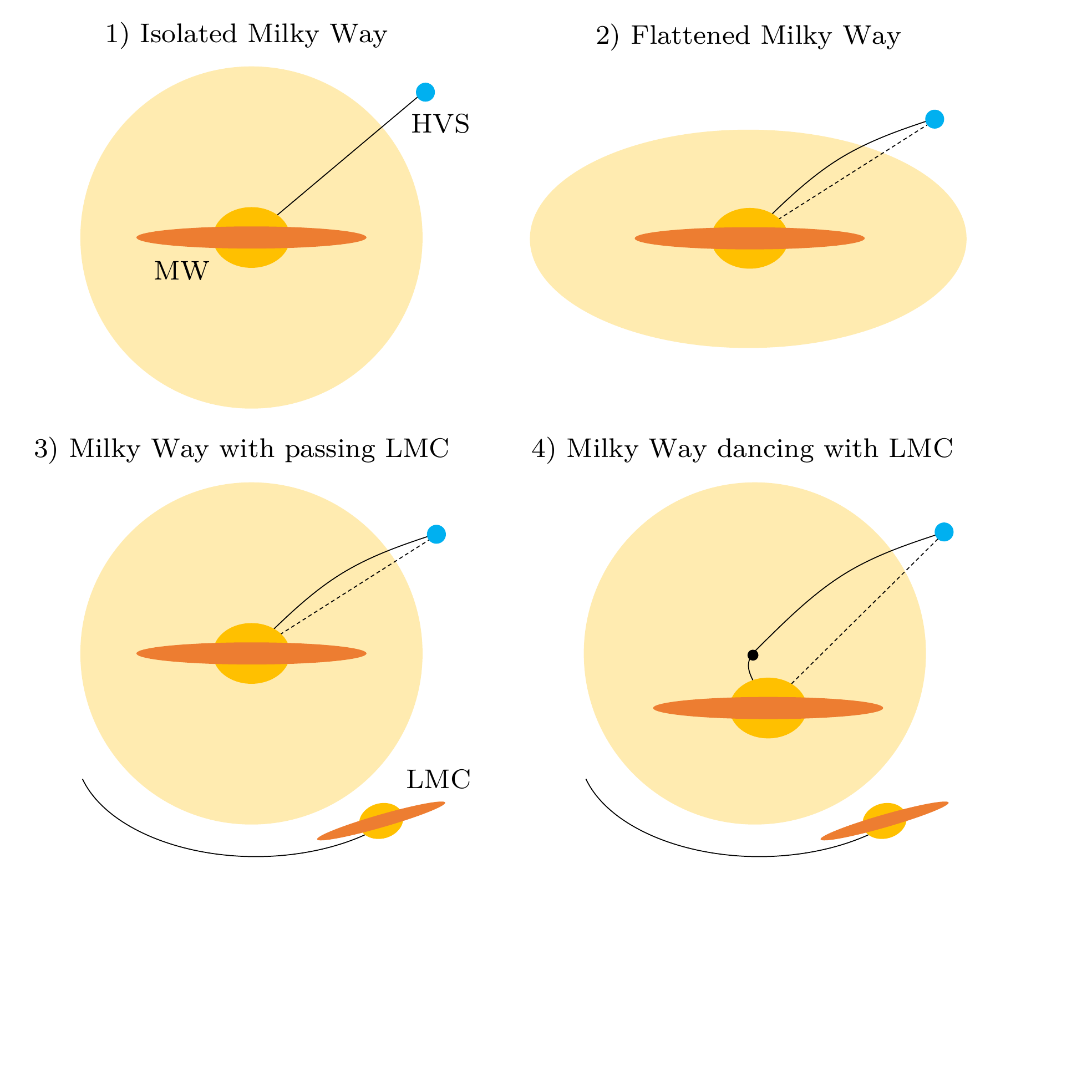}
	\caption{Schematic diagram of the four scenarios considered in this work. The Milky Way and LMC are shown as orange ellipses and an escaping hypervelocity star as a blue circle, with the solid black lines indicating their past trajectories. The dashed black line connects the hypervelocity star to the present-day location of the Galactic centre. \textbf{1)} A hypervelocity star ejected from the centre of an isolated Milky Way will escape along a straight line trajectory (neglecting the potential of the bulge, bar and disk). \textbf{2)} If the dark matter halo of the Milky Way is flattened into an oblate spheroid, then the trajectory of the hypervelocity star will be deflected towards the plane. \textbf{3)} The Isolated Milky Way but with the fly-past of the LMC included and the Milky Way fixed-in-place. The hypervelocity star is deflected towards the LMC. \textbf{4)} Same as above, but with the Milky Way free to move in response to the LMC potential. The motion of the Milky Way since the time of ejection changes the relative position and velocity of the hypervelocity star that we measure today.}
	\label{fig:diagram}
\end{figure*}

The fastest stars in the Milky Way are moving fast enough to escape from the Galactic gravitational field. These stars have a speed above the escape speed at their location and so are known as \textit{hypervelocity stars}. While the hypervelocity neutron stars \citep{Lyne1994,Arzoumanian2002}, white dwarfs \citep{Shen2018,Raddi2018,Raddi2019} and sub-dwarfs \citep{Hirsch2005,Geier2015} are thought to have their origin as the runaway remnants or former companions of supernovae, the non-compact nature of the main sequence hypervelocity stars precludes that as their origin. The main sequence hypervelocity stars -- most of which are early-type stars far from the Galactic disk -- are instead thought to have been gravitationally slingshotted during the tidal disruption of a binary star by the supermassive black hole Sgr A* at the Galactic centre, in a phenomenon known as the \citet{Hills1988} mechanism. The first evidence of the Hills mechanism came with the discovery of a $3\;\mathrm{M}_{\odot}$ main sequence hypervelocity star travelling at over $700\;\mathrm{km}\;\mathrm{s}^{-1}$ more than $100\;\mathrm{kpc}$ away \citep{Brown2005}. However, neither that star nor any of the forty or so main sequence hypervelocity stars discovered since \citep{Brown2015} have a precise enough distance or proper motion to be tracked definitively back to the Galactic centre. The association of the hypervelocity stars with the Galactic centre was thus significantly bolstered by the recent discovery of S5-HVS1 by \citet{Koposov2019}: a $2.3\;\mathrm{M}_{\odot}$ star travelling at $1700\;\mathrm{km}\;\mathrm{s}^{-1}$ that tracks back exactly to the Galactic centre. 

S5-HVS1 will traverse a further $2\;\mathrm{Mpc}$ over the remainder of its $1\;\mathrm{Gyr}$ lifespan. The trajectory of S5-HVS1 will be deflected away from a straight line by the non-spherical and time-evolving components of the gravitational potential it traverses: by the bulge and disk, then by the dark matter halo and the satellite galaxies of the Milky Way, and finally by Andromeda and the other members of the Local Group. \citet{Gnedin2005} showed that a measurement of the deflection of a hypervelocity star's trajectory could be used to infer the shape of the Galactic dark matter halo, and \citet{Kenyon2018} investigated the deflection that would be caused by the Milky Way disk and the Large Magellanic Cloud (LMC). The missing ingredient from these previous works is that the Milky Way itself experiences the time-evolving gravitational field of the Local Group. Consequently, the Milky Way moves and changes our perspective on the geometry of an escaping straight-line trajectory. The hypervelocity stars were ejected from a Milky Way that was in a different location and moving at a different velocity, and thus cannot be assumed to have begun their journey at the present day location of the Galactic centre.

The time-evolving potential that the Milky Way is experiencing is currently dominated by the LMC, the most massive of the Galactic satellites. The mass of the LMC has been subject to much debate, but recent estimates have converged on a total mass slightly larger than $10^{11}\;\mathrm{M}_{\odot}$. \citet{Kallivayalil2013} argued that the LMC must be at least $10^{11}\;\mathrm{M}_{\odot}$ for it to have held onto the SMC for a reasonable fraction of a Hubble time, whilst \citet{Penarrubia2016} modified the local Hubble flow timing argument to account for the LMC and found a likely mass of $2.5\times10^{11}\;\mathrm{M}_{\odot}$. \citet{Erkal2019} showed that the evident perturbation in the Orphan stellar stream required an LMC mass of $1.4\times10^{11}\;\mathrm{M}_{\odot}$, while \citet{Erkal2019b} argued that a further six dwarf satellite galaxies fell in with the LMC and thus that the mass must be greater than $1.2\times10^{11}\;\mathrm{M}_{\odot}$. The potential disturbance of the Milky Way due to the LMC's passage is therefore substantial: the LMC is one-tenth the mass of the Galaxy, is only a handful of scale-radii away \citep{McMillan2011}, and is flying past at $321\;\mathrm{km}\;\mathrm{s}^{-1}$ \citep{Kallivayalil2013}. \edit{\cite{Gomez:2015} showed that this will induce a substantial reflex motion in the Milky Way.} The innermost regions of the Milky Way \edit{(within $\sim30$ kpc)} will respond adiabatically as the Galaxy is pulled down towards the LMC, whereas the outer stellar halo will feel a weaker gravitational force and so will fall behind, resulting in a net $20\;\mathrm{km}\;\mathrm{s}^{-1}$ upwards motion of the halo relative to the Milky Way\edit{, depending on the mass of the LMC} \citep[as shown in simulation by][]{Garavito2019,Petersen:2020,Erkal:2020}. The stars and dark matter near the LMC will be directly affected, with \citet{Garavito2019} predicting that this material will form a wake trailing the LMC, which has now been \edit{tentatively identified as the Pisces Overdensity \citep{Watkins:2009}} by \citet{Belokurov2019} using RR Lyrae stars in \textit{Gaia} DR2 and Pan-STARRS1 \edit{along with a spectroscopic sample of Blue Horizontal Branch stars and Blue Stragglers}.

The LMC is such a substantial galaxy that it itself has ejected hypervelocity stars. The $8\;\mathrm{M}_{\odot}$ hypervelocity star HE 0437-5439 was discovered by \citet{Edelmann2005}, who conjectured that this star may originate in the LMC due to their small separation of $18\;\mathrm{kpc}$. \citet{Gualandris2007} affirmed this association and suggested that an intermediate black hole in the LMC could have caused the ejection, and final proof came with proper motions from \textit{Gaia} DR2 which showed that HE 0437-5439 was ejected from the LMC $21.1^{+6.1}_{-4.6}\;\mathrm{Myr}$ ago at $870^{+69}_{-66}\;\mathrm{km}\;\mathrm{s}^{-1}$ \citep{Erkal2019c}. Theoretical work by \citet{Boubert2016} and \citet{Boubert2017} has shown that the ejection of stars from the LMC by either the Hills mechanism or supernovae in massive binary stars could imply the existence of hundreds or thousands of hypervelocity stars associated with the LMC. The remainder of this work focuses specifically on the hypervelocity stars ejected from the Milky Way, but we conjecture that the two populations could prove to be complementary probes.

The ratio of the distance and velocity of the LMC relative to the Milky Way gives a timescale of roughly $150\;\mathrm{Myr}$, comparable to the flight times of the hypervelocity stars on the outskirts of the Milky Way, making this perturbation relevant to their escape trajectories. There are two related effects:
\begin{enumerate}
    \item The escape trajectory of hypervelocity stars is deflected by the gravitational field of the LMC.
    \item The centre of the Milky Way itself moves in response to the LMC, changing the position and velocities of the hypervelocity stars relative to the Galactic centre.
\end{enumerate}
The first of these effects was investigated by \citet{Kenyon2018}, but the investigation of the second of these effects in this work is entirely novel. As we will show, these two effects are vital to include and are of comparable magnitude, as exaggeratedly illustrated in Fig. \ref{fig:diagram}. 

In this work, we simulate the ejection of hypervelocity stars from the Galactic centre over the last \edit{$1\;\mathrm{Gyr}$} from four different Milky Ways: an isolated Milky Way (traditional), an isolated Milky Way with a flattened dark matter halo (considered by \citealp{Gnedin2005}), a fixed-in-place Milky Way with a passing LMC \citep[equivalent to][]{Kenyon2018}, and a Milky Way that freely moves in response to the LMC (novel in this work). We investigate the varied consequences of this Galactic motion, finding that many of the hypervelocity stars do not track back to the Galactic centre and that the pattern of deflection on the sky is radically changed. We conclude that if we wish to use the deflection of the hypervelocity stars to trace the shape of the Galactic dark matter halo, then we must acknowledge that the Milky Way moves.

\section{Methodology}
\label{sec:simulation}
We illustrate the influence of the LMC on our interpretation of the hypervelocity stars by simulating their production over the last \edit{$1\;\mathrm{Gyr}$} and integrating their orbits in a selection of potentials. This paper is merely a demonstration that accounting for the reflex motion of the Milky Way is necessary, and thus in Sec. \ref{sec:initial} we use a simple analytic prescription to generate a population of escaping hypervelocity stars. In Sec. \ref{sec:nbody} we give the simple analytic potentials used for the Milky Way and LMC, and describe the injection and orbit integration of the hypervelocity stars.

\subsection{Initial conditions for hypervelocity stars}
\label{sec:initial}

An analytic prescription for ejection by the Hills mechanism is given by \citet[also described by \citealp{Kenyon2014}]{Bromley2006}, which predicts a probability distribution of ejection velocities of hypervelocity stars as a function of the total mass $M_1+M_2$ and separation $a_{\mathrm{bin}}$ of the intercepting binary, the mass of the black hole $M_{\mathrm{bh}}$, and the radius of closest approach $r_{\mathrm{close}}$ of the binary to the black hole. We sample these properties from standard distributions to construct a population of single and binary stars in the Galactic centre. We sample the primary mass $M_1$ from the \citet{Kroupa2001} initial mass function\footnote{We note that previous work has found an extremely top-heavy initial mass function for the young stars in the Galactic centre \citep{Bartko2010}, i.e. the population which may produce a substantial fraction of the hypervelocity stars, and future work should account for this.},
\begin{equation}
N(M_1)\propto
\begin{cases}
M_1^{-0.3}, & \mathrm{if}\ 0.01<M_1/\mathrm{M}_{\odot}<0.08, \\
M_1^{-1.3}, & \mathrm{if}\ 0.08<M_1/\mathrm{M}_{\odot}<0.5, \\
M_1^{-2.3}, & \mathrm{if}\ 0.5<M_1/\mathrm{M}_{\odot}<80.0, \\
0, & \mathrm{otherwise,}
\end{cases}
\end{equation}
and assign companions in a probabilistic way based on the $M_1$-dependent, analytic, empirically-motivated binary fraction described by \citet{Arenou2010},
\begin{equation}
F_{\mathrm{bin}}(M_1)=0.8388\tanh(0.079+0.688M_1).
\end{equation}
The single stars are discarded and for the binaries we sample the mass ratio $q$ from a uniform distribution over the range $0.1\;\mathrm{M}_{\odot}/M_1<q<1$. The period $P$ is sampled from the log-normal distribution of \citet{Duquennoy1991}, i.e. a normal distribution
in $\log_{10}(P/\mathrm{days})$ with a mean of 4.8 and a standard
deviation of 2.3, truncated to lie between -2.0 and 12.0. The period is readily converted to the separation of the binary $a_{\mathrm{bin}}$ using Kepler's Third Law.

\citet{Bromley2006} note that the more massive of the two stars is more likely to be ejected and that the ejection of both stars is rare. For simplicity we assume that the ejection of either star is equally likely and is independent of the masses of the two stars. We choose to consider only $2-4\;\mathrm{M}_{\odot}$ hypervelocity stars because this is roughly the mass of the known B-type HVSs on the outskirts of the Milky Way, and so discarded any hypervelocity stars outside this mass range to avoid integrating the orbits of stars that were unlikely to survive to the outskirts (high mass) or to be too faint to see once they arrived there (low mass).

If a binary is to produce a hypervelocity star that is visible today, then that progenitor binary must last long enough to encounter Sgr A* and the star which is ejected must survive until present day. We use an analytic equation for the main sequence lifetime $\tau_{\mathrm{MS}}$ \citep{Hurley2000} which gives the two edge cases $\tau_{\mathrm{MS}}(2\;\mathrm{M}_{\odot})=1.16\;\mathrm{Gyr}$ and $\tau_{\mathrm{MS}}(4\;\mathrm{M}_{\odot})=0.17\;\mathrm{Gyr}$. The time that the binary is born $\tau_{\mathrm{born}}$ is sampled uniformly between $1.16\;\mathrm{Gyr}$ ago and the present. We assume that the likelihood of a binary encountering Sgr A* is directly proportional to how long the binary survives, and so the time of encounter $\tau_{\mathrm{enc}}$ is drawn uniformly from the range $(\tau_{\mathrm{born}},\tau_{\mathrm{born}}+1.16\;\mathrm{Gyr})$. If $\tau_{\mathrm{enc}}$ is after the present day then we discard that binary.

For those binaries which encounter Sgr A* before present day, we follow \citet{Kenyon2014} in sampling the distance of closest approach $r_{\mathrm{close}}$ from the distribution
\begin{equation}
P(r_{\mathrm{close}})\propto r_{\mathrm{close}},
\end{equation}
where $r_{\mathrm{close}}$ is constrained to lie between $1\;\mathrm{and}\;1000\;\mathrm{AU}$. The inner edge is chosen to avoid collisions between the two stars and Sgr A*, and the outer edge prevents us from sampling too many ejected stars which will not escape the Galactic centre.

\citet{Kenyon2014} states that the ejection velocity from the encounter of a binary with a massive black hole can be approximated as a random draw from the Normal distribution
\begin{equation}
\label{eq:pvej}
P(v_{\text{ej}})\;\text{d}v_{\text{ej}} \propto \exp{\left(-\frac{( v_{\text{ej}} - v_{\text{ej,H}} )^2}{2\sigma_v^2}\right)} \;\text{d} v_{\text{ej}},
\end{equation}
where $\sigma_v= \frac{1}{5} v_{\text{ej,H}}$ and $v_{\text{ej,H}}$ is given by
\begin{align}
\label{eq:vejH}
v_{\mathrm{ej,H}}=&1760\left(\frac{a_{\mathrm{bin}}}{0.1 \; \mathrm{AU}}\right)^{-1/2}\left(\frac{M_1+M_2}{2 \; \mathrm{M}_{\odot}}\right)^{1/3}\nonumber  \\
&\times\left(\frac{M_{\mathrm{bh}}}{3.5 \times 10^6 \; \mathrm{M}_{\odot}}\right)^{1/6} f_{\mathrm{R}}\; \mathrm{km}\;\mathrm{s}^{-1}.
\end{align}
The quantity $f_{\mathrm{R}}$ is a normalisation factor and is given by
\begin{align}
\label{eq:fr}
f_{\mathrm{R}}&=0.774+(0.0204+(-6.23 \times 10^{-4} + (7.62\times 10^{-6} \nonumber \\
&+(-4.24\times 10^{-8}+8.62\times 10^{-11}D) D)D)D)D,
\end{align}
where $D$ encodes the dependency on the binary separation $a_{\mathrm{bin}}$ and closest approach $r_{\mathrm{close}}$ through
\begin{equation}
\label{eq:d}
D=\left(\frac{r_{\mathrm{close}}}{a_{\mathrm{bin}}}\right)\left[\frac{2M_{\mathrm{bh}}}{10^6(M_1+M_2)}\right]^{-1/3}.
\end{equation}
We fix $M_{\mathrm{BH}}$ at the mass of Sgr A*, which we take to be $4\times 10^6\;\mathrm{M}_{\odot}$. The probability of the ejection $P_{\mathrm{ej}}$ of one of the stars depends on the binary phase and the orientation of the binary orbit relative to the black hole \citep{Bromley2006},
\begin{equation}
\label{eq:pej}
P_{\mathrm{ej}}=1-D/175.
\end{equation}
We sample a number from $U(0,1)$ and discard any system where this number is greater than $P_{\mathrm{ej}}$. We choose one star of the binary to eject with equal probability and with the velocity of the ejected component being given by \citet{Bromley2006} as
\begin{equation}
v_1=v_{\mathrm{ej}}\left(\frac{2M_{2}}{M_1+M_2}\right)^{1/2},\;v_2=v_{\mathrm{ej}}\left(\frac{2M_{1}}{M_1+M_2}\right)^{1/2}.
\end{equation}
The final constraints we apply are that the ejected star must have a mass $M_{\mathrm{ej}}$ that satisfies $2<M_{\mathrm{ej}}<4\;\mathrm{M}_{\odot}$, that it must survive on the main sequence until present day, and that the ejection velocity is at least $600\;\mathrm{km}\;\mathrm{s}^{-1}$. This velocity cut approximately corresponds to a star that after ejection from the Galactic centre will arrive at $1\;\mathrm{kpc}$ with zero velocity, and was chosen to minimise the number of stars we simulate which will remain deep within the MW potential.

We iterate the procedure above to generate $10^7$ stars ejected by the Hills mechanism operating at the centre of the Milky Way. The 75\% percentile of the resulting ejection velocity distribution occurs at $808\;\mathrm{km}\;\mathrm{s}^{-1}$, the 95\% at $1131\;\mathrm{km}\;\mathrm{s}^{-1}$, the 99\% at $1566\;\mathrm{km}\;\mathrm{s}^{-1}$, and the 99.9\% at $2459\;\mathrm{km}\;\mathrm{s}^{-1}$.

\begin{figure}
	\includegraphics[scale=0.49,trim = 10.5mm 0mm 0mm 12mm, clip]{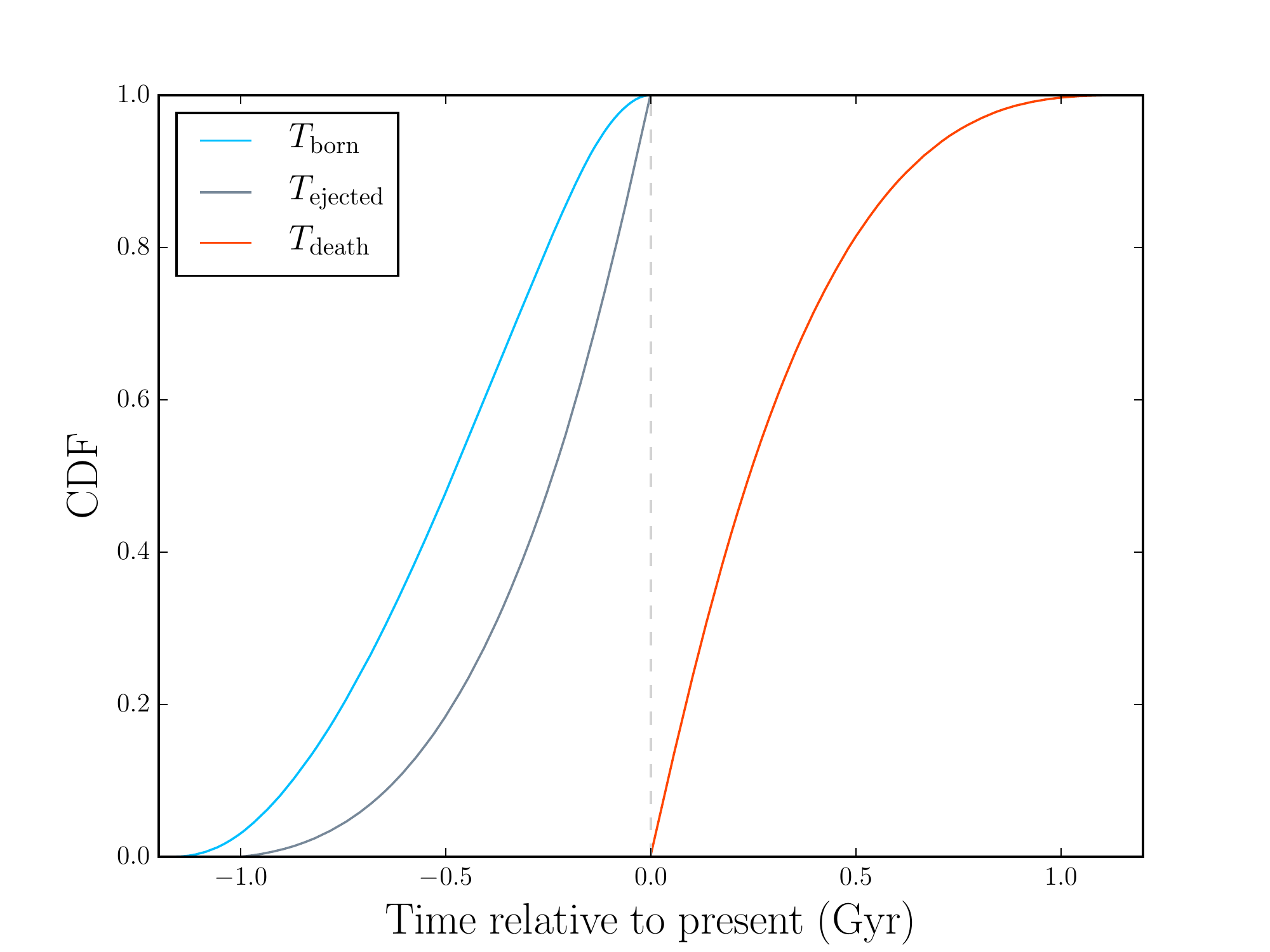}
	\caption{The cumulative density of the time of birth, ejection and death for our simulated sample of hypervelocity stars. This sample was selected such that they were ejected before and survived until present day.}
	\label{fig:threetimescales}
\end{figure}

In Fig. \ref{fig:threetimescales} we show the cumulative distribution functions of the three timescales involved in the ejection of stars by the Hills mechanism; the time that the binary stars is born $T_{\mathrm{born}}$, the time at which the binary star interacts with Sgr A* and one star is ejected $T_{\mathrm{ejected}}$, and the time at which the ejected star reaches the end of the main-sequence $T_{\mathrm{death}}$. Note that we have only included stars if they are ejected prior to and survive until present day.

\subsection{Details of the orbit integrations}
\label{sec:nbody}

We ran four simulations to isolate the two different effects that the fly-by of the LMC has on the escape trajectories of hypervelocity stars, as illustrated in Fig. \ref{fig:diagram}:
\begin{enumerate}
	\item \textbf{Isolated Milky Way} -- The gravitational potential is only that of the Milky Way. The only non-spherical components of the potential are the bulge and disk.
	\item \textbf{Flattened Milky Way} -- Same as the Isolated Milky Way but with an extremely oblate dark matter halo.
	\item \textbf{Milky Way with passing LMC} -- The gravitational potential of the Large Magellanic Cloud is included and that dwarf galaxy flies past in response to the Galactic potential, but the Milky Way is held in place. The hypervelocity stars are deflected by the time-evolving potential of the LMC.
	\item \textbf{Milky Way dancing with LMC} -- The Milky Way and the Large Magellanic Cloud move in response to their mutual gravitational attraction. The location of the hypervelocity stars on the sky changes due to the motion of the Milky Way, and the hypervelocity stars experience the time evolving potential from both galaxies moving.
\end{enumerate}

The Milky Way potential we used in the Isolated Milky Way, Passing LMC and Dancing with LMC simulations was \edit{similar to \texttt{MWPotential2014} from \cite{Bovy2015} and was} composed of a \citet{Hernquist1990} bulge
\begin{equation}
\label{eq:potentialb}
\phi_{\mathrm{b}}(r) = -GM_{\mathrm{b}}/(r+a_{\mathrm{b}}),
\end{equation}
with
$M_{\mathrm{b}}=5\times10^9\; \mathrm{M}_{\odot}$ and
$a_{\mathrm{b}} = 500 \; \mathrm{pc}$, a \citet{Miyamoto1975} disk
\begin{equation}
\label{eq:potentiald}
\phi_{\mathrm{d}}(R,z) = -GM_{\mathrm{d}}/\sqrt{R^2+\left[a_{\mathrm{d}}+\left( z^2+b_{\mathrm{d}}^2 \right)^{1/2}\right]},
\end{equation}
with
$M_{\mathrm{d}}=6.8\times10^{10} \; \mathrm{M}_{\odot}$,
$a_{\mathrm{d}}=3000 \; \mathrm{pc}$ and $b_{\mathrm{d}}=280 \;
\mathrm{pc}$, and a \citet{Navarro1997} dark matter halo
\begin{equation}
\label{eq:potentialh}
\phi_{\mathrm{h}}(r) = -\frac{GM_{\mathrm{h}}}{r} \frac{\ln{\left(1+r/r_{\mathrm{h}}\right)}}{\ln(1+\mathrm{c})-\frac{\mathrm{c}}{1+\mathrm{c}}},
\end{equation}
with $M_{\mathrm{h}} =
8\times 10^{11} \; \mathrm{M}_{\odot}$, $r_{\mathrm{h}} = 16 \;\mathrm{kpc}$, and a concentration of $\mathrm{c}=15.3$. In the Flattened Milky Way simulation the bulge and disk were the same, but the dark matter NFW halo was flattened to have an axis ratio $c/a = 0.5$. As we discuss in Sec. \ref{sec:implications}, this flattening is extreme and so the deflection of the hypervelocity stars caused by this dark matter halo should be considered to be a limiting case. The Large Magellanic Cloud potential was a Hernquist potential with $M_{\mathrm{b}}=1.5\times10^{11}\; \mathrm{M}_{\odot}$ and $a_{\mathrm{b}} = 17.13 \; \mathrm{kpc}$. This is motivated by the results of \cite{Erkal2019} and matches the observed rotation curve of the LMC at 8.7 kpc \citep{vandermarel_2014}. \edit{We included dynamical friction on the LMC from the Milky Way using the prescription from \cite{Jethwa:2016}. For the present day distance, radial velocity, and proper motions of the LMC we used the values from \cite{pietrzynski_lmc_dist, vandermarel_lmc_rv,Kallivayalil2013} respectively.}

\edit{In order to account for the reflex motion of the Milky Way, we treat the LMC and Milky Way as individual particles sourcing their respective potentials. Thus we treat them as rigid profiles throughout and do not account for their expected tidal deformation. The Milky Way and LMC are first integrated backwards from their present day positions for 1 Gyr. The integration is done using a leapfrog kick-drift-kick integrator. During the forward integration, the hypervelocity stars are injected as tracer particles at their respective ejection times as described in Sec. \ref{sec:initial}. They are initialised at a distance of 1.4 pc from the location of the Galactic centre at their time of ejection with a randomly oriented but initially radial trajectory. Note that for the flattened Milky Way halo, we use \textsc{galpot} \citep{Dehnen:1998} to evaluate the forces but integrate the orbits with our own integrator.}

After integrating the orbits until present day, we generated synthetic astrometric, photometric and spectroscopic measurements for the stars by using \textsc{parsec} isochrones to convert the initial mass into a $G$ band magnitude and assuming the Sun is located at $(R,z)=(8.3,0.027)\;\mathrm{kpc}$ and moving at $(v_R,v_\phi,v_z)=(-11.1,232.24,7.25)\;\mathrm{km}\;\mathrm{s}^{-1}$ relative to the present day position and velocity of the Milky Way \citep[this is the frame used by pre-v4.0 \textsc{astropy}, based on measurements from][]{Chen2001,Gillessen2009,Schonrich2010,Bovy2015}.

In the discussion in Sec. \ref{sec:discussion} we will focus on the two first-order effects of the deflection of the hypervelocity stars by the LMC and the change in perspective due to the reflex motion of the Milky Way, but we note that our simulation accounts for the second order effect of the motion of the Milky Way changing the potential that both the LMC and the hypervelocity stars experience, which causes further changes in the trajectories of the LMC and hypervelocity stars, which causes the Milky Way to change its motion in response, and so on. 


\section{Results and discussion}
\label{sec:discussion}

We illustrate how the LMC complicates the landscape in which we must study the hypervelocity stars from four perspectives. First, we show that hypervelocity stars that have travelled for more than $100\;\mathrm{Myr}$ in our realistic `Milky Way dancing with LMC' simulation do not track back to the Galactic centre of an isolated Milky Way, and which could lead to us inferring the wrong place of ejection. Second, we map the deflections of the hypervelocity stars away from a straight-line trajectory in each simulation, and show that the LMC is the dominant deflector. Third, we show that future proper motion measurements combining \textit{Gaia} with the proposed \textit{Gaia}NIR mission would be sufficiently precise to probe both the deflection of the hypervelocity stars by the LMC and the change in perspective caused by the Milky Way moving towards the LMC. This motion is substantial: we found that over the last \edit{$1\;\mathrm{Gyr}$} the Milky Way has moved $4.3\;\mathrm{kpc}$ with a velocity change of $29.6\;\mathrm{km}\;\mathrm{s}^{-1}$ in response to the LMC. Fourth, we discuss the implications for inferring the shape of the Milky Way using hypervelocity stars, because we will make an extremely biased inference of the dark matter halo triaxiality if we neglect the interaction of the Milky Way and LMC.

\subsection{HVSs do not track back to the Galactic centre}

 \begin{figure}
	\includegraphics[width=\columnwidth,trim = 0mm 0mm 0mm 0mm, clip]{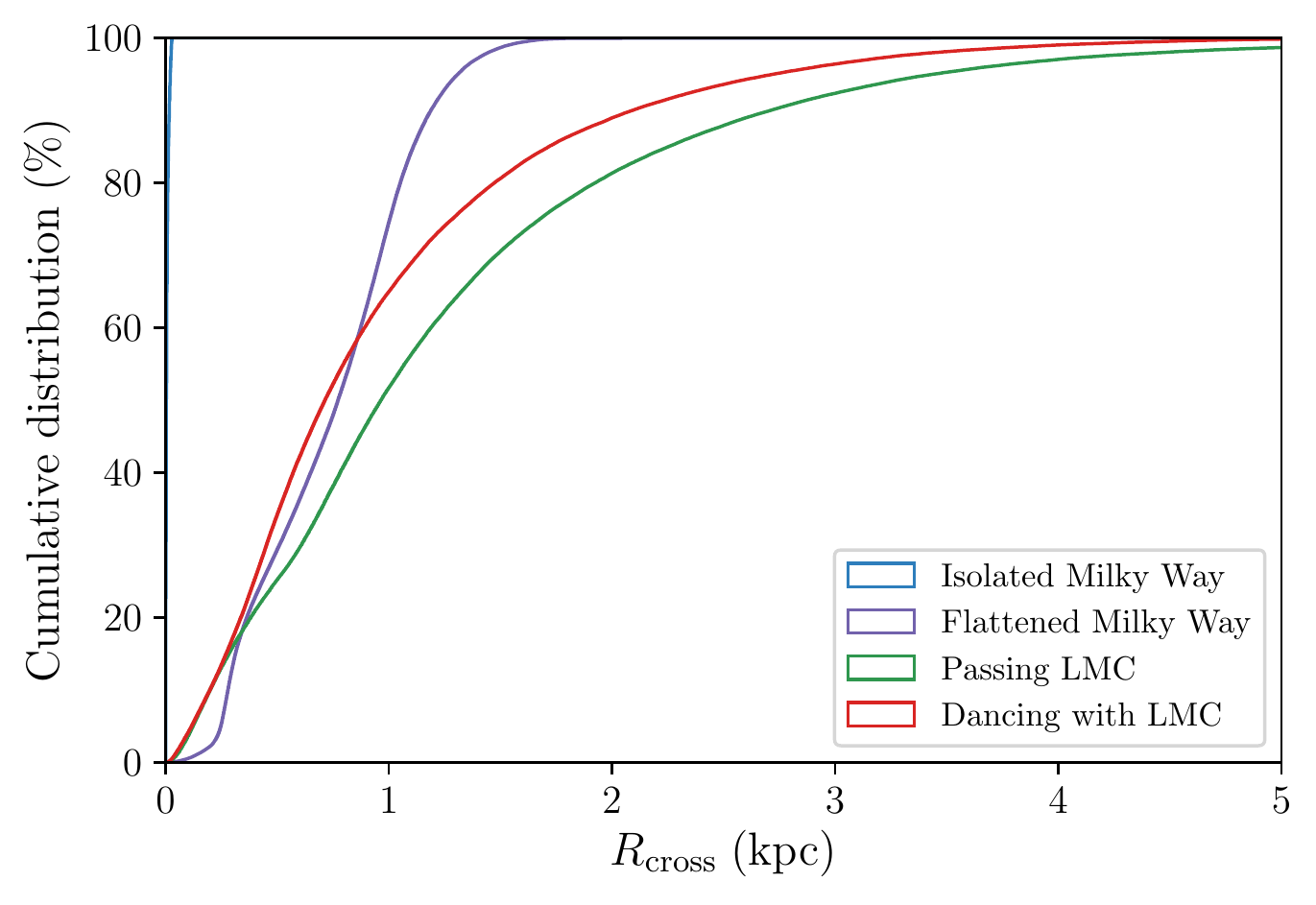}
	\caption{The simulated stars were tracked backwards in time to their last crossing of the Galactic plane in a potential that only includes the Milky Way. The cumulative distributions of the distance of this crossing from the Galactic centre $R_{\mathrm{cross}}$ gives an indication of how consistent each population is with having been ejected from the Galactic centre, noting that we only show here those stars that are escaping the Milky Way and track back to the plane more than $100\;\mathrm{Myr}$ ago. The line corresponding to the stars from the `Isolated Milky Way' potential is barely visible on the lefthand edge of this plot, because these stars are being integrated backwards in the same potential in which they were integrated forwards and so track back exactly to the Galactic centre (modulo small numerical errors). Stars from the `Flattened Milky Way' simulation all cross within $2\;\mathrm{kpc}$, while neglecting the existence of the LMC causes as many as 10\% of the stars from the other two simulations to cross more than $2\;\mathrm{kpc}$ from the Galactic centre.}
	\label{fig:cross}
\end{figure}

A common approach when investigating the origins of a hypervelocity star is to track the orbit of the star backwards in time until it last crossed the plane of the Milky Way's disk. If the hypervelocity star was ejected from the Galactic centre then that crossing location should be exactly the Galactic centre, subject to us having correctly modelled the gravitational potential that the hypervelocity star has experienced. Ignoring the potential of the LMC or the motion of the Milky Way will bias the crossing location, causing us to wrongly infer that the hypervelocity star did not originate in the Galactic centre. To quantify this bias, we integrated the orbits of the stars from each of the four simulations backwards in time assuming the Isolated Milky Way potential. We made the sample somewhat realistic by restricting our consideration to the stars in each simulation that at present day are brighter than $G=20$ and have Galactocentric velocities above the escape speed from a static Milky Way. In Fig. \ref{fig:cross} we show the cumulative density of the crossing radius $R_{\mathrm{cross}}$ in each of the four simulations, restricted to those stars that cross the Isolated Milky Way plane at least $100\;\mathrm{Myr}$ ago. That some of the stars from the `Isolated Milky Way' simulation do not integrate back exactly to the Galactic centre ($R_{\mathrm{cross}}=0\;\mathrm{kpc}$) indicates that there is some small level of numerical noise. Nevertheless, all of these stars do integrate back within $21\;\mathrm{pc}$, which is in extreme contrast with the other three simulations. In the `Flattened Milky Way' simulation, the $(5,95)\%$ region of $R_{\mathrm{cross}}$ is $(0.25,1.31)\;\mathrm{kpc}$, indicating that the missing deflection of the oblate dark matter halo causes very few of the hypervelocity stars to track back near to the Galactic centre, but all still track back to within a couple kiloparsecs. The  $(5,95)\%$ regions in the `Passing LMC' and `Dancing with LMC' simulations are $(0.12,3.44)\;\mathrm{kpc}$ and $(0.12,2.73)\;\mathrm{kpc}$ respectively. The reason that many of the stars integrate back close to the Galactic centre in the two simulations with the LMC (as apposed to the `Flattened Milky Way' simulation) is that the Milky Way used in these simulations is the same as that in the `Isolated Milky Way' potential in which we back-track the stars. However, more than 10\% track back outside $2\;\mathrm{kpc}$ in both simulations, which is further than could be explained by the Milky Way being triaxial.

Ignoring the LMC will cause you to infer that the hypervelocity stars did not come from the Galactic centre, even if you measure the position and velocity of the star precisely. We note that the uncertainties in the crossing locations of the hypervelocity stars on the outskirts of the Milky Way are usually of order a few to a few tens of kiloparsecs \citep{Brown2015a} and thus that this effect will likely only become detectable in future.

\subsection{Deflection from radial trajectory}

\begin{figure*}
	\includegraphics[width=\linewidth,trim = 0mm 0mm 0mm 0mm, clip]{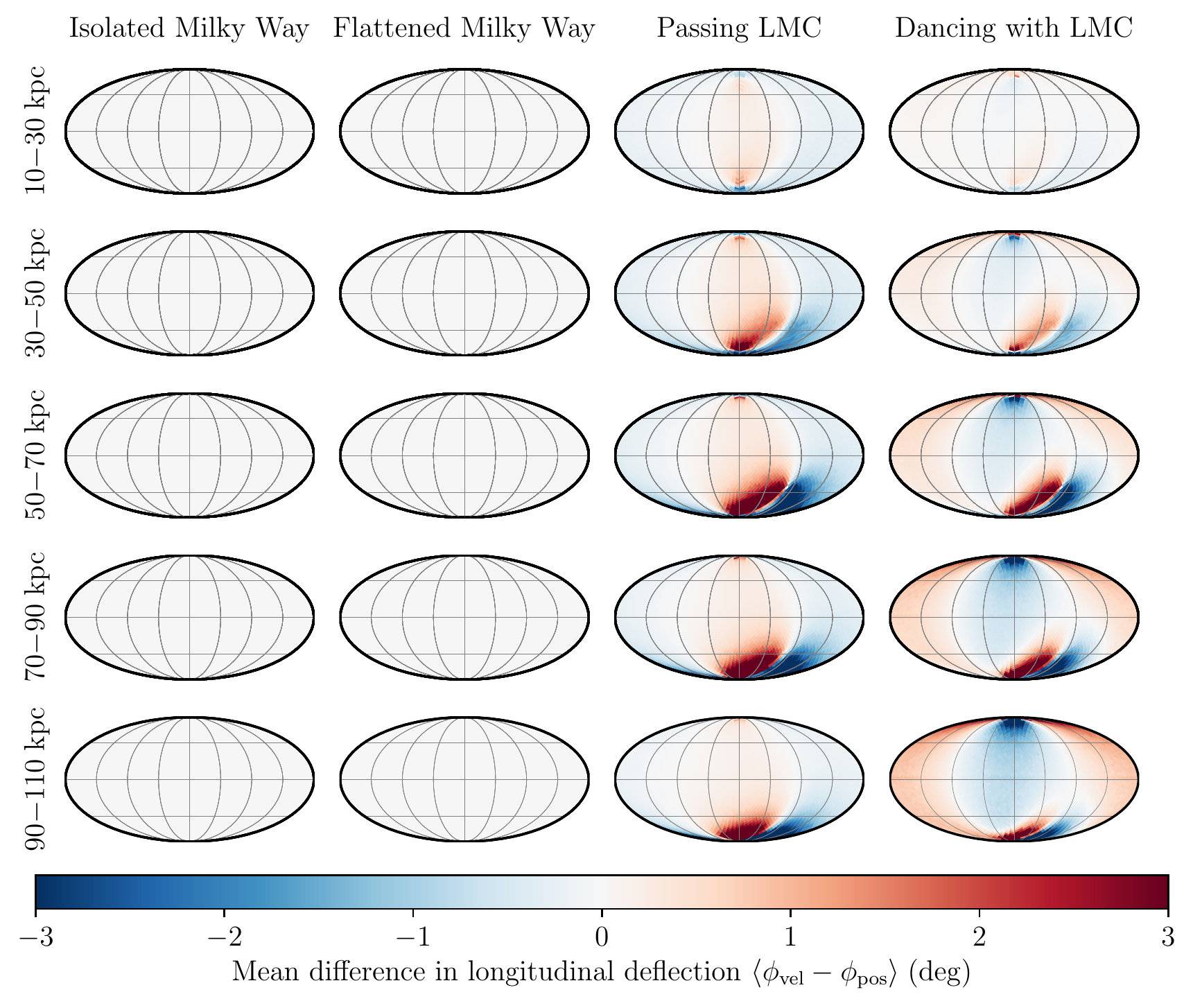}
	\caption{Galactic maps of the mean deflection in the longitudinal direction of stars in each of the four simulations, split by their Galactocentric radius. We assume the Milky Way potential to be axisymmetric and so the only longitudinal deflection is from the LMC. The simulated stars are deflected towards the LMC, but the additional apparent deflection from the motion of the Milky Way towards the LMC in the `Dancing with LMC' simulations causes the total deflection to flip sign for stars which are far from the LMC.}
	\label{fig:longitude}
\end{figure*}

\begin{figure*}
	\includegraphics[width=\linewidth,trim = 0mm 0mm 0mm 0mm, clip]{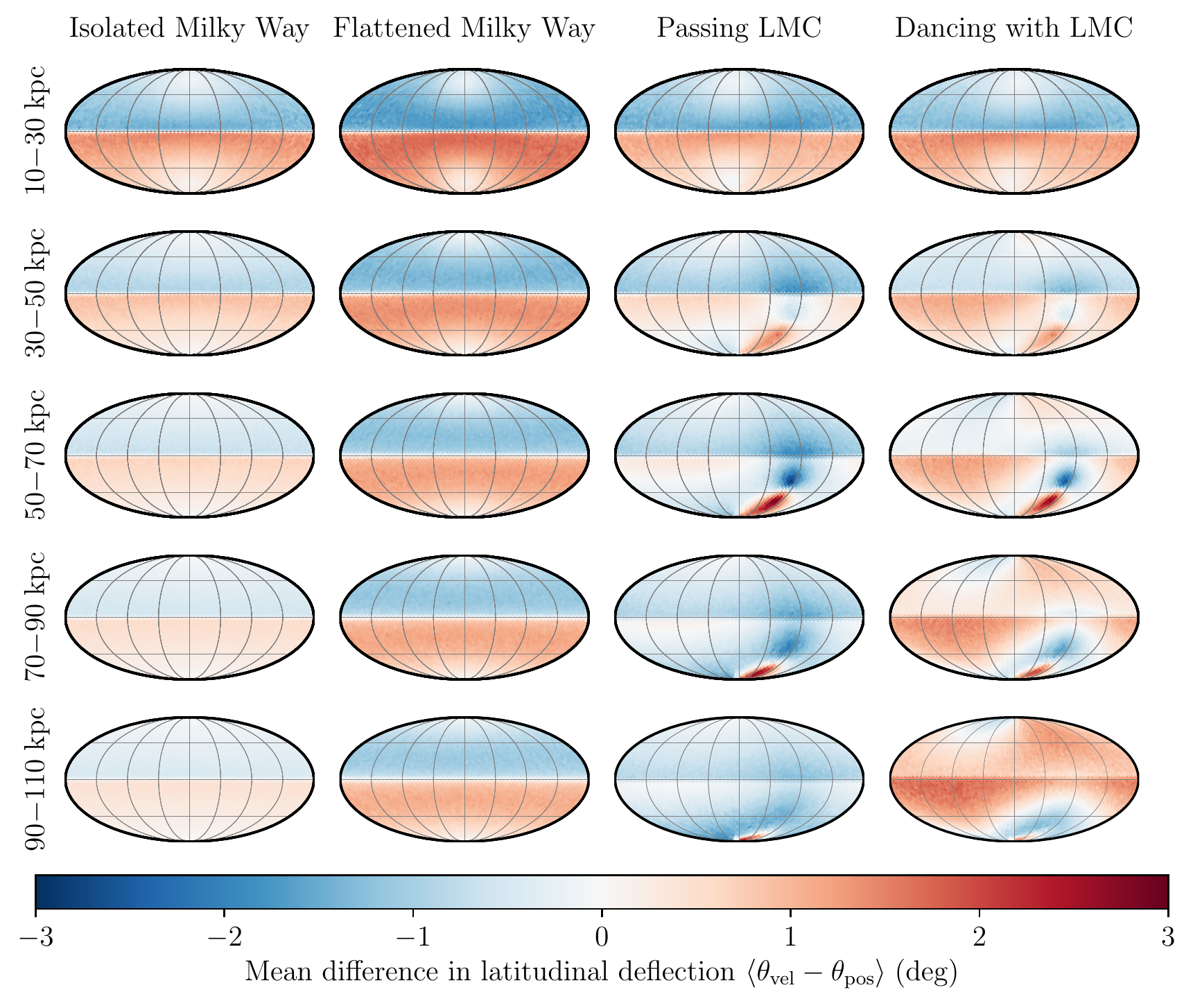}
	\caption{Same as Fig. \ref{fig:longitude}, but for the latitudinal deflections. The Milky Way's bulge and disk deflect stars towards the plane, with this deflection being enhanced by the oblate dark matter halo in the `Flattened Milky Way' case, but this deflection is sub-dominant to those caused by the LMC and the reflex of the Milky Way at large Galactocentric radii.}
	\label{fig:latitude}
\end{figure*}

If the Milky Way were perfectly spherical and isolated, then a hypervelocity star ejected from the Galactic centre would escape along an exactly radial trajectory. Any non-sphericity and time-evolution in the potential that the hypervelocity star has traversed causes that trajectory to deviate, which we can quantify through the angles between the star's position and velocity vectors.

From the simulations run in Sec. \ref{sec:simulation} we have the Galactocentric positions $(x,y,x)$ and velocities $(v_x,v_y,v_z)$ at present-day of all the stars ejected from the Galactic centre. We restrict ourselves to stars whose total velocity $v=\sqrt{v_x^2+v_y^2+v_z^2}$ is greater than the escape speed at the current location of the star $v_{\mathrm{esc}}(r)$, where $r=\sqrt{x^2+y^2+z^2}$ and we calculate the escape speed using the `Isolated Milky Way' potential. We made this cut because stars not moving fast enough to escape orbit the Milky Way and so have pseudo-randomised angles between their position and velocity vectors. We define the position $(\phi_{\mathrm{pos}},\theta_{\mathrm{pos}})$ and velocity $(\phi_{\mathrm{vel}},\theta_{\mathrm{vel}})$ angles by
\begin{equation}
    \tan{\phi_{\mathrm{pos}}}\equiv \frac{y}{x},\;\sin{\theta_{\mathrm{pos}}} \equiv \frac{z}{r},\;\tan{\phi_{\mathrm{vel}}}\equiv\frac{v_y}{v_x},\;\sin{\theta_{\mathrm{vel}}}\equiv\frac{v_z}{v}.
\end{equation}
We calculated these for every escaping star in each simulation and then found the average longitudinal $\langle \phi_{\mathrm{vel}}-\phi_{\mathrm{pos}} \rangle$ and latitudinal $\langle \theta_{\mathrm{vel}}-\theta_{\mathrm{pos}} \rangle$ deflections in bins in radius $r$ and in $\textsc{nside}=16$ HEALPix \citep{Gorski2005} pixels on the sky, which we show in Figs. \ref{fig:longitude} and \ref{fig:latitude} respectively. A trajectory is radial if and only if  $\phi_{\mathrm{vel}}-\phi_{\mathrm{pos}}=0$ and $\theta_{\mathrm{vel}}-\theta_{\mathrm{pos}}=0$.

We note that these deflection angles can be equivalently expressed in terms of the orbital angular momentum $\vec{L}$ and the angle between the position and velocity vectors $\psi$, for instance,
\begin{equation}
    \tan{\left(\phi_{\mathrm{vel}}-\phi_{\mathrm{pos}}\right)} = \frac{L_z}{rv\cos{\psi}}.
\end{equation}
A hypervelocity star escaping along a perfectly radial trajectory has zero angular momentum; the deflections shown in Figs. \ref{fig:longitude} and \ref{fig:latitude} represent the change of the angular momentum away from zero.

There is no longitudinal deflection (Fig. \ref{fig:longitude}) in the `Isolated Milky Way' or `Flattened Milky Way' cases because the potentials are axisymmetric. There are large longitudinal deflections at all radii in the `Passing LMC' simulation because stars are pulled towards the LMC's orbit, causing the sky to be split by which side of the LMC's path the stars are on. In the `Dancing with LMC' simulation the stars nearest the LMC show a similar pattern of deflection, but the large scale pattern of longitudinal deflection across the rest of the sky is flipped, because the Milky Way is closer to the LMC than the hypervelocity stars at these locations, and so the motion of the hypervelocity stars towards the LMC is more than offset by the greater motion of the Milky Way towards the LMC. We note that there are only weak longitudinal deflections of the hypervelocity stars nearest to the Milky Way in the `Dancing with LMC' simulation, which demonstrates the unphysical nature of the `Passing LMC' simulation; the Milky Way and the hypervelocity stars at that radius are feeling almost the same gravitational force from the LMC, but only the hypervelocity stars are moving in response to it.

The latitudinal deflection (Fig. \ref{fig:latitude}) in the `Isolated Milky Way' case is due to the disk and bulge pulling stars down towards the plane, and this deflection is enhanced by the oblate dark matter halo in the `Flattened Milky Way' simulation. These deflections decrease in magnitude with distance due to a geometric effect where the velocity vector of an escaping star becomes more radial with increasing distance from the Galaxy, assuming that the non-spherical force acting on the star has become negligible. The gravitational pull from the disk or halo is greater than that from the LMC for stars closer than $30\;\mathrm{kpc}$, and so the top row of Fig. \ref{fig:latitude} is fairly unchanged by the inclusion of the LMC. There is a stark disagreement between the `Passing LMC' and `Dancing with LMC' simulations farther out: at $90{-}110\;\mathrm{kpc}$ there are locations on the sky where the mean deflection is $1.5\;\mathrm{deg}$ downwards for the former and $1.5\;\mathrm{deg}$ upwards for the latter. Over more than 90\% of the sky, the predicted deflection in the `Isolated Milky Way' simulation is closer to the true `Dancing with LMC' deflections than the `Passing LMC' approximation. At large radii the deflection in the `Flattened Milky Way' simulation is sub-dominant to the deflection from the LMC.

\edit{We note that both Figs. \ref{fig:longitude} and \ref{fig:latitude} show broad similarity with Fig. 1 of \citet{Erkal:2020}, who simulated the proper motion deflections of stars in the Milky Way halo by an infalling LMC. The difference in our case is that hypervelocity stars pass the LMC so rapidly that their trajectories are not sufficiently deflected for an LMC-trailing wake to form.}

We conclude that interpreting the angle between the Galactocentric position and velocity vectors of the distant hypervelocity stars will require us to account for the motion of the Milky Way in response to the LMC.

\subsection{Proper motion prediction error}
\label{sec:propermotion}
\begin{figure*}
	\includegraphics[width=\linewidth,trim = 0mm 0mm 0mm 0mm, clip]{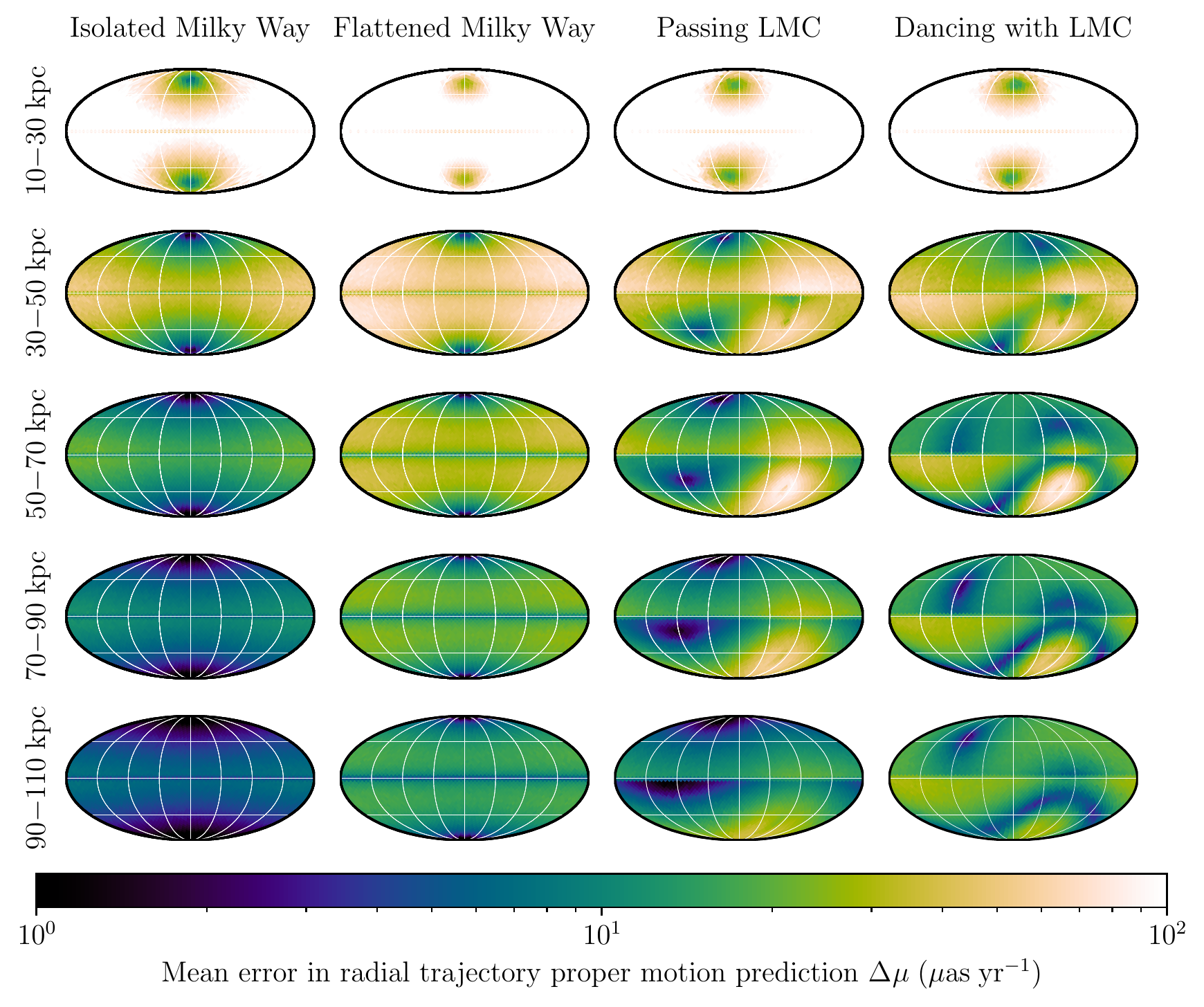}
	\caption{If we make the simplistic assumption that the simulated stars are escaping from the Milky Way on perfectly radial trajectories, then we can predict the proper motion of the stars from their measured longitude, latitude, distance and radial velocity. This figure shows maps of the mean total error between this simple predicted proper motion and the true proper motion in pixels on the sky, split by Galactocentric radius. At large radii the deflection caused by the LMC and reflex of the Milky Way causes proper motion prediction errors of a few tens of $\mu\mathrm{as}\;\mathrm{yr}^{-1}$, much larger than the few $\mu\mathrm{as}\;\mathrm{yr}^{-1}$ error caused by the deflection from the Galactic bulge and disk.}
	\label{fig:propermotion}
\end{figure*}

\begin{figure}
	\includegraphics[width=\linewidth,trim = 0mm 0mm 0mm 0mm, clip]{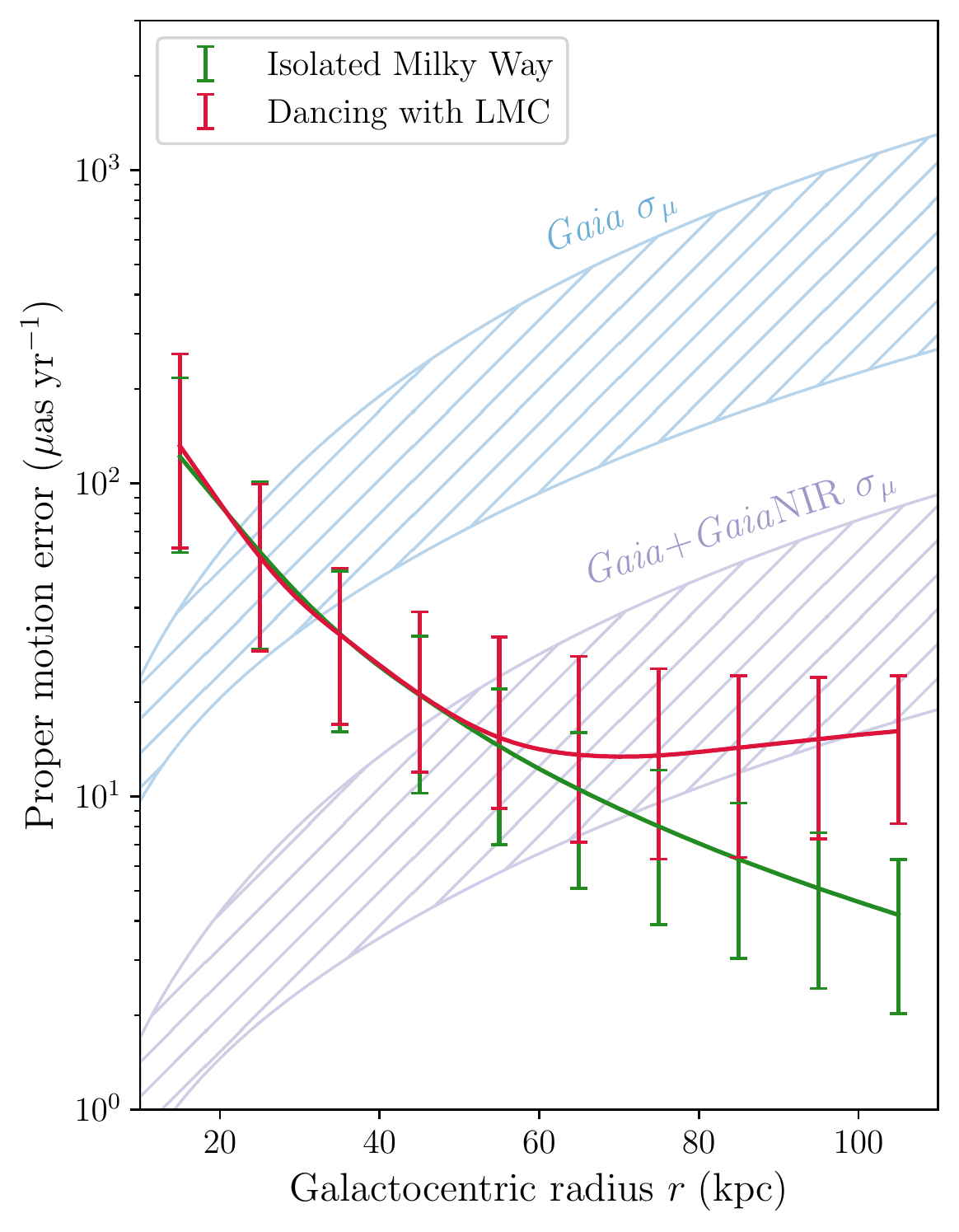}
	\caption{The radial trajectory proper motion prediction error (shown here in $10\;\mathrm{kpc}$ bins of Galactocentric radius) is dominated at large radii by the joint deflection caused by the LMC and the response of the Milky Way to the LMC. Overplotted are predicted proper motion uncertainty envelopes for $2\;\mathrm{M}_{\odot}$ (top) and $4\;\mathrm{M}_{\odot}$ (bottom) stars as measured by \textit{Gaia} and by the combination of \textit{Gaia} with a proposed near-infrared successor \textit{Gaia}NIR launching in twenty years time.}
	\label{fig:propermotionerrors}
\end{figure}

If a star is at Galactocentric radius $r$ and moving with velocity $v$, then the angles $(\phi_{\mathrm{pos}},\theta_{\mathrm{pos}})$ and $(\phi_{\mathrm{vel}},\theta_{\mathrm{vel}})$ allow us to compute the Galactocentric positions and velocities,
\begin{equation}
\begin{bmatrix} x \\ y \\ z \end{bmatrix}
= \begin{bmatrix} r \cos{\phi_{\mathrm{pos}}}\cos{\theta_{\mathrm{pos}}} \\ 
r \sin{\phi_{\mathrm{pos}}}\cos{\theta_{\mathrm{pos}}} \\  
r \sin{\theta_{\mathrm{pos}}} \end{bmatrix}, \quad
\begin{bmatrix} v_x \\ v_y \\ v_z \end{bmatrix}
= \begin{bmatrix} v \cos{\phi_{\mathrm{vel}}}\cos{\theta_{\mathrm{vel}}} \\ 
v \sin{\phi_{\mathrm{vel}}}\cos{\theta_{\mathrm{vel}}} \\  
v \sin{\theta_{\mathrm{vel}}} \end{bmatrix}.
\end{equation}
If the star is moving perfectly radially then $\phi_{\mathrm{pos}}=\phi_{\mathrm{vel}}=\phi$ and $\theta_{\mathrm{pos}}=\theta_{\mathrm{vel}}=\theta$, and thus knowing any four of the six kinematic quantities (or any transform of those to a different frame) is sufficient to deduce the other two. Historically, the easiest kinematic quantities to measure were the position $(l,b)$ and the heliocentric radial velocity $v_r$ and distance $d$, and these could then be used to estimate the proper motion and total velocity of a hypervelocity star under the assumption that it is escaping on a perfectly radial trajectory.

We will investigate in this section how badly wrong this predicted proper motion is in each of the three simulations, assuming that the position $(l,b)$ and the heliocentric radial velocity $v_r$ and distance $d$ are perfectly known. For each of the stars in the three simulations we compute the component of the velocity $v_r$ that lies along the vector with Galactic angles $(l,b)$ from the Sun:
\begin{equation}
    v_r = -v_x\cos{l}\cos{b}+v_y\sin{l}\cos{b}+v_z\sin{b}.
\end{equation}
We then found the Galactocentric total velocity $\tilde{v}$ and velocity vector $(\tilde{v}_x,\tilde{v}_y,\tilde{v}_z)$ of a star with that same heliocentric radial velocity that was on a perfectly radial trajectory:
\begin{equation}
    \tilde{v} = \frac{-v_x\cos{l}\cos{b}+v_y\sin{l}\cos{b}+v_z\sin{b}}{-\cos{\phi}\cos{\theta}\cos{l}\cos{b}+\sin{\phi}\cos{\theta}\sin{l}\cos{b}+\sin{\theta}\sin{b}}.
\end{equation}
Both sets of angles $(\phi,\theta)$ and $(l,b)$ can be computed from the Galactocentric position of the star $(x,y,z)$ together with the Solar position $(R_{\odot},z_{\odot})$. Finally, we predicted the true proper motions $(\mu_{\alpha\ast},\mu_{\delta})$ of each star and the predicted proper motions $(\tilde{\mu}_{\alpha\ast},\tilde{\mu}_{\delta})$ assuming the star is on a perfectly radial trajectory. We defined the radial trajectory proper motion prediction error $\Delta\mu$ to be
\begin{equation}
    \Delta\mu = \sqrt{(\tilde{\mu}_{\alpha\ast}-\mu_{\alpha\ast})^2+(\tilde{\mu}_{\delta}-\mu_{\delta})^2},
\end{equation}
and show maps of this error in bins in radius $r$ and in $\textsc{nside}=16$ HEALPix pixels on the sky in Fig. \ref{fig:propermotion}. 

Assuming a perfectly radial trajectory is a poor assumption in the inner regions where the Galactic disk and bulge are dominant, which is apparent in the large proper motion prediction errors in the top row of Fig. \ref{fig:propermotion}. In the `Isolated Milky Way' case this assumption improves with distance and the residual proper motion prediction error is less than a few $\mu\mathrm{as}\;\mathrm{yr}^{-1}$ in the $90{-}110\;\mathrm{kpc}$ bin. The `Flattened Milky Way' case shows a similar pattern of error but with a larger amplitude due to the additional deflection from the oblate dark matter halo. In the other two simulations, the deflection from the LMC is already the dominant source of error across large parts of the sky at radii greater than $30\;\mathrm{kpc}$, but the pattern of proper motion errors is different in both cases. We note the `ring' around the LMC that appears in the `Dancing with LMC' simulation, which is where the downwards motion of the hypervelocity stars towards the LMC is equal to the reflex motion of the Milky Way. Understanding the proper motions of the distant hypervelocity stars will require us to account for the LMC deflecting both the hypervelocity stars and the Milky Way.

We have shown that the proper motions of the hypervelocity stars on the outskirts of the Milky Way will be perturbed by the LMC, but the resulting proper motion prediction errors shown in Fig. \ref{fig:propermotion} are only a few tens of $\mu\mathrm{as}\;\mathrm{yr}^{-1}$. To quantify the size of the signal, we binned the escaping stars in the `Isolated Milky Way' and `Dancing with LMC' simulations in $10\;\mathrm{kpc}$ bins and calculated the 16\%, 50\% and 84\% percentiles of the proper motion prediction error $\Delta\mu$. We show these percentiles in Fig. \ref{fig:propermotionerrors}. At radii closer than $50\;\mathrm{kpc}$ the median $\Delta\mu$ are coincident, but further out the curves begin to deviate and by the $100{-}110\;\mathrm{kpc}$ bin the $(16,84)\%$ regions are entirely disjoint. However, to distinguish these two curves will require proper motion measurements with precisions of roughly $10\;\mu\mathrm{as}\;\mathrm{yr}^{-1}$.

The most precise proper motion measurements which will be available in the near term are those from the \textit{Gaia} space mission \citep{Gaia2016}. While \textit{Gaia} proper motion measurements\footnote{\url{https://www.cosmos.esa.int/web/gaia/science-performance}} will reach this precision for bright stars $V<12$ after the initial five year mission, the hypervelocity stars at these distances are fainter than $V>17$. We estimated the proper motion errors $\sigma_{\mu,\mathrm{5 yr}}$ we should expect for $2\;\mathrm{M}_{\odot}$ and $4\;\mathrm{M}_{\odot}$ stars at each Galactocentric radius by the end of the initial five year \textit{Gaia} mission by estimating $V$ and $I$ using PARSEC isochrones and applying the formulae on the \textit{Gaia} science performance webpage. The \textit{Gaia} mission is likely to be extended by up to five years and hence we further scaled the proper motions to get the expected ten year end-of-mission proper motion errors $\sigma_{\mu}=\sigma_{\mu,\mathrm{5 yr}}/2^{3/2}$, which we show in Fig. \ref{fig:propermotionerrors}. We conclude that the proper motions measured by \textit{Gaia} will not be precise enough to detect any deflection by the LMC. However, there is hope for the future in the form of the proposed \textit{Gaia}NIR successor mission \citep{Hobbs2016}, which would repeat the \textit{Gaia} mission in the near-infrared. The likely twenty year gap between the two missions would result in extremely precise proper motions. If we assume that the two missions have similar end-of-mission proper motion uncertainties then the improvement is roughly a factor of 14 over the ten year \textit{Gaia} measurements, and we indicate these improved proper motion uncertainties in Fig. \ref{fig:propermotionerrors}. These combined $\mathit{Gaia}+\mathit{Gaia}\mathrm{NIR}$ proper motions would allow us to detect the deflection due to the LMC for $4\;\mathrm{M}_{\odot}$ stars between $60{-}90\;\mathrm{kpc}$. In future it will be possible to use these deflections to infer the change in position and velocity of the Milky Way in response to the gravitational pull of the LMC.

We have opted not to separate the proper motion prediction error into right ascension and declination for brevity, but we note that the deflection caused by the Galactic disk and LMC will deflect stars in different directions, and thus the two deflections will be easier to disentangle in the 2D proper motion space.

\subsection{Implications for measuring the triaxiality of the Milky Way}
\label{sec:implications}

A majority of the Milky Way's mass is in the dark matter halo and it is possible that that halo is elongated, flattened or even triaxial. \citet{Prada2019} considered the shapes of dark matter halos in the Auriga cosmological simulations and found that baryons act to make halos oblate spheroids, where the minor-to-major axis ratios $c/a$ were in the range $0.6{-}0.9$. \citet{Chua2019} reached a similar conclusion after analysing the Illustris simulation, finding that the median minor-to-major axis ratio was $\langle c/a \rangle \approx 0.7$ for Milky Way-like halo masses ($\leq 10^{12.5} \mathrm{M}_{\odot}$). Simulations of galactic mergers which include central supermassive black holes also find moderately oblate spheroids \citep{Bortolas2018}, where the central black holes act to moderate the oblateness. The axis ratio $c/a=0.5$ considered in our `Flattened Milky Way' simulation can thus be considered an extreme case and so illustrates the maximum deflection of the hypervelocity stars that can be expected from the halo.

\citet{Gnedin2005} investigated the deflection of hypervelocity stars in a variety of possible halo shapes, and found that a proper motion precision of $10\;\mu\mathrm{as}\;\mathrm{yr}^{-1}$ would be required to infer the shape of the halo from the trajectories of the hypervelocity stars, which agrees with our findings in Fig. \ref{fig:propermotion}. We note that this is only slightly smaller than the deflections that we predict to be caused by the LMC, which we find will cause changes in their proper motions of about $10-30\;\mu\mathrm{as}\;\mathrm{yr}^{-1}$. We can therefore conclude that the pattern of hypervelocity star deflection across the sky will likely reflect a composition of the dominant LMC deflection with a minor deflection by the oblate dark matter halo.

We leave a detailed investigation of the combined deflections due to these two sources to a future work, nearer to the time when proper motion precisions will make this a practicable measurement, but note that using the hypervelocity stars to infer the shape of the Milky Way's dark matter halo will be biased unless we account for the LMC moving the Milky Way and hypervelocity stars.

\section{Conclusions}
\label{sec:conclusions}

The hypervelocity stars escape from the centre of the Milky Way out to the edges of the Local Group, but are deflected from a straight line trajectory by the non-spherical and time-evolving gravitational field that they traverse. Previous works have comprehensively investigated the deflections caused by the Milky Way and the Large Magellanic Cloud (LMC), but have neglected to account for the Milky Way itself moving in response to the LMC. A consequence of this motion is that the hypervelocity stars we see on the outskirts of the Milky Way today were ejected from where the Milky Way centre was hundreds of millions of years ago. This change in perspective causes large apparent deflections in the trajectories of the hypervelocity stars which are of the same order as the deflections caused by the non-spherical components of the potential.

While proper motion measurements from \textit{Gaia} will not reach the $10\;\mu\mathrm{as}\;\mathrm{yr}^{-1}$ precision necessary to probe these deflections in the hypervelocity stars' trajectories, this precision will be reached if the near-infrared successor mission, \textit{Gaia}NIR, is launched in twenty years and reaches the same astrometric precision as \textit{Gaia} for these stars. At that point, it would be possible to use the hypervelocity stars' deflections as an independent probe of the shape of the Milky Way's dark matter halo, the mass of the Large Magellanic Cloud, and of the dance of these two galaxies about each other.

\section*{Acknowledgements}
DB thanks Magdalen College for his fellowship and the Rudolf Peierls Centre for Theoretical Physics for providing office space and travel funds. This work has made use of data from the European Space Agency (ESA) mission
{\it Gaia} (\url{https://www.cosmos.esa.int/gaia}), processed by the {\it Gaia}
Data Processing and Analysis Consortium (DPAC,
\url{https://www.cosmos.esa.int/web/gaia/dpac/consortium}). Funding for the DPAC
has been provided by national institutions, in particular the institutions
participating in the {\it Gaia} Multilateral Agreement.




\bibliographystyle{mnras}
\bibliography{references} 




%
%


\bsp	
\label{lastpage}
\end{document}